\documentclass[%
 reprint,
 amsmath,amssymb,
 aps,
]{revtex4-2}

\usepackage{graphicx}
\usepackage{dcolumn}
\usepackage{bm}
\usepackage{hyperref}

 \hypersetup{
        colorlinks=true,
        linkcolor=blue,
        filecolor=blue,
        urlcolor=blue,
        citecolor=blue
    }

\usepackage{amsmath}
\usepackage{xcolor}
\begin{document}

\preprint{APS/123-QED}

\title{Out-of-plane angle resolved second harmonic Hall analysis in perpendicular magnetic anisotropy systems}

\author{Akanksha Chouhan}
\thanks{Corresponding author Email: ecakanksha@ee.iitb.ac.in}%
\author{Abhishek Erram}
\author{Ashwin A. Tulapurkar}

\affiliation{Department of Electrical Engineering, Indian Institute of Technology Bombay, Mumbai 400076, India}

\date{\today}

\begin{abstract}
Spin orbit torques (SOT) are used to manipulate magnetization of ferromagnets in heavy metal / ferromagnetic bilayers and  thus are technologically relevant from magnetization switching perspective. This necessitates development of methods which give an insight into SOT behaviour and enable SOT efficiency estimation. In this article, we demonstrate an experimental approach of out-of-plane (OOP) angle resolved Second Harmonic Hall (SHH) measurement, in Pt/Co and Ta/CoFeB perpendicular magnetic anisotropy (PMA) systems, for damping-like and field-like SOT efficiency estimation. This method reveals a unique anomalous field-like torque dependent on magnetization direction in Ta/CoFeB system. The damping-like and field-like SOT efficiencies are extracted by solving LLGS equation in the low frequency limit of magnetic susceptibility. Along with SHH measurements, we also experimentally demonstrate anomalous Hall effect (AHE) based spin-torque ferromagnetic resonance (STFMR) technique for SOT efficiency quantification in PMA systems.
\end{abstract}

\maketitle

\section{\label{sec:level1}Introduction}
Spintronic memory devices rely on magnetization switching caused by spin orbit torques. In the heavy metal (HM)/ferromagnetic (FM) bilayers in spintronic devices, an rf charge current flowing through HM undergoes spin orbit coupling resulting in oscillating spin current which is absorbed by the FM. This spin current exerts torque on the FM magnetization and causes magnetization manipulation which could result in magnetization switching. The spin orbit torques are classified into damping-like and field-like torques. The spin current that is generated in HM is not fully transferred to FM because of the spin backflow and spin scattering at the interface \cite{pai2015dependence}. Hence, efficiency metrics need to be defined for damping-like and field-like torques, which are called as damping-like spin orbit torque efficiency ($\xi_{DL}$) and field-like spin orbit torque efficiency ($\xi_{FL}$). Accurate estimation of these spin orbit torque efficiencies is necessary for proper design of competitive spintronic memory and logic devices.

Various characterization techniques exist for estimation of SOT efficiency out of which spin torque ferromagnetic resonance (STFMR) and Second Harmonic Hall (SHH) are the prominent ones. The STFMR, specifically the anisotropic magnetoresistance (AMR) based STFMR \cite{liu2011spin,tulapurkar2005spin,bose2017sensitive} is conducive to in-plane magnetic anisotropy ferromagnet (FM)/heavy metal (HM) bilayer systems. This method does not lend itself to perpendicular magnetic anisotropy (PMA) FM/HM bilayers as AMR is directly related to FM thickness and PMA FM layers are very thin \cite{wei2020characterization}. Research on  an alternate Spin Hall Magnetoresistance (SMR) based STFMR measurement in PMA systems has been reported in ref \cite{wei2020characterization,he2018spin,he2016spin}. However, the SHH method \cite{pi2010tilting,kim2013layer,hayashi2014quantitative,garello2013symmetry,pai2014enhancement,yun2017accurate} remains the mainstay for estimation of SOT efficiency in PMA FM/HM bilayer systems. The prevalent SHH method involves estimation of SOT efficiencies through low range field sweep \cite{kim2013layer,garello2013symmetry,pai2014enhancement}as well as through in-plane angle sweep \cite{avci2014interplay,roschewsky2019spin,dutta2021interfacial}. Small angle out of plane angle sweep has been reported in ref\cite{shirokura2021angle, qiu2014angular}.

In the SHH measurement technique, a small frequency ac conduction current flowing through the heavy metal layer generates an oscillating spin current which when injected in PMA ferromagnetic layer causes SOT modulation thereby inducing magnetization oscillations. These magnetization oscillations result in modulation of anomalous Hall effect (AHE) and planar Hall effect (PHE) resistances, which is detectable in the second harmonic Hall voltage signal. Thus, from the SHH voltage signal, damping-like ($\xi_{DL}$) and field-like ($\xi_{FL}$) SOT efficiencies can be estimated.

In the SHH measurement technique as Hall voltage is being recorded, input current and recorded voltage are transverse to each other. The field sweep way of recording SHH voltage involves injecting a low frequency ac current in $\hat{x}$ direction and a transverse voltage measurement, once for an external magnetic field sweep in $\hat{x}$ direction (longitudinal case) and next for field sweep in $\hat{y}$ direction (transverse case). In the in-plane (azimuthal) angle sweep way of recording SHH voltage, the external magnetic field is rotated in the plane of the sample.

In this article, we demonstrate an SHH approach wherein SHH voltage is recorded for a full $360^\circ$ out-of-plane (polar) angle scan of applied external magnetic field. A comparison of OOP angle sweep in two PMA systems viz. Pt/Co and Ta/CoFeB is  presented. Along with SOT efficiency estimation, this method offers an insight into the behaviour of SOT with respect to the magnetization direction, which is lacking in other SHH based approaches. Specifically, our data on the Ta/CoFeB system indicates a surprising feature that the field-like spin-orbit torque term depends on the direction of magnetization. 
Further we demonstrate the equivalence of measurement of second harmonic Hall voltage and dc voltage. 
The estimation of SOT efficiencies from the recorded SHH data is done by taking a different route wherein the low frequency limit of magnetic susceptibility is exploited to solve LLGS equation. Additionally, AHE based STFMR method is presented for SOT efficiency estimation in PMA systems. The results of Pt/CO PMA system are shown.


\section{OOP Angle Resolved Harmonic Hall Analysis: Experimental Results}
Harmonic Hall measurements were performed on Hall bar devices of dimension 10$\times$20 $\mu$$m^2$ patterned on Si/SiO2 substrate by optical lithography followed by PMA stack deposition using magnetron sputtering. We studied two PMA systems - i) Ta(1)/Pt(5)/Co(1)/MgO(1.5)/Ta(1.5) and ii) Ta(5)/CoFeB(1.3)/MgO(1.5)/Ta(1.5), where numbers in bracket represent thickness in nm. All metal layers in the stacks were deposited using DC magnetron sputtering at a rate of 1nm/min and the dielectric MgO was deposited through RF sputtering at a rate of 0.2nm/min. 

\begin{figure}[!b]
   \centering
   \vspace{0.0in}
    \includegraphics[width=3in, keepaspectratio]{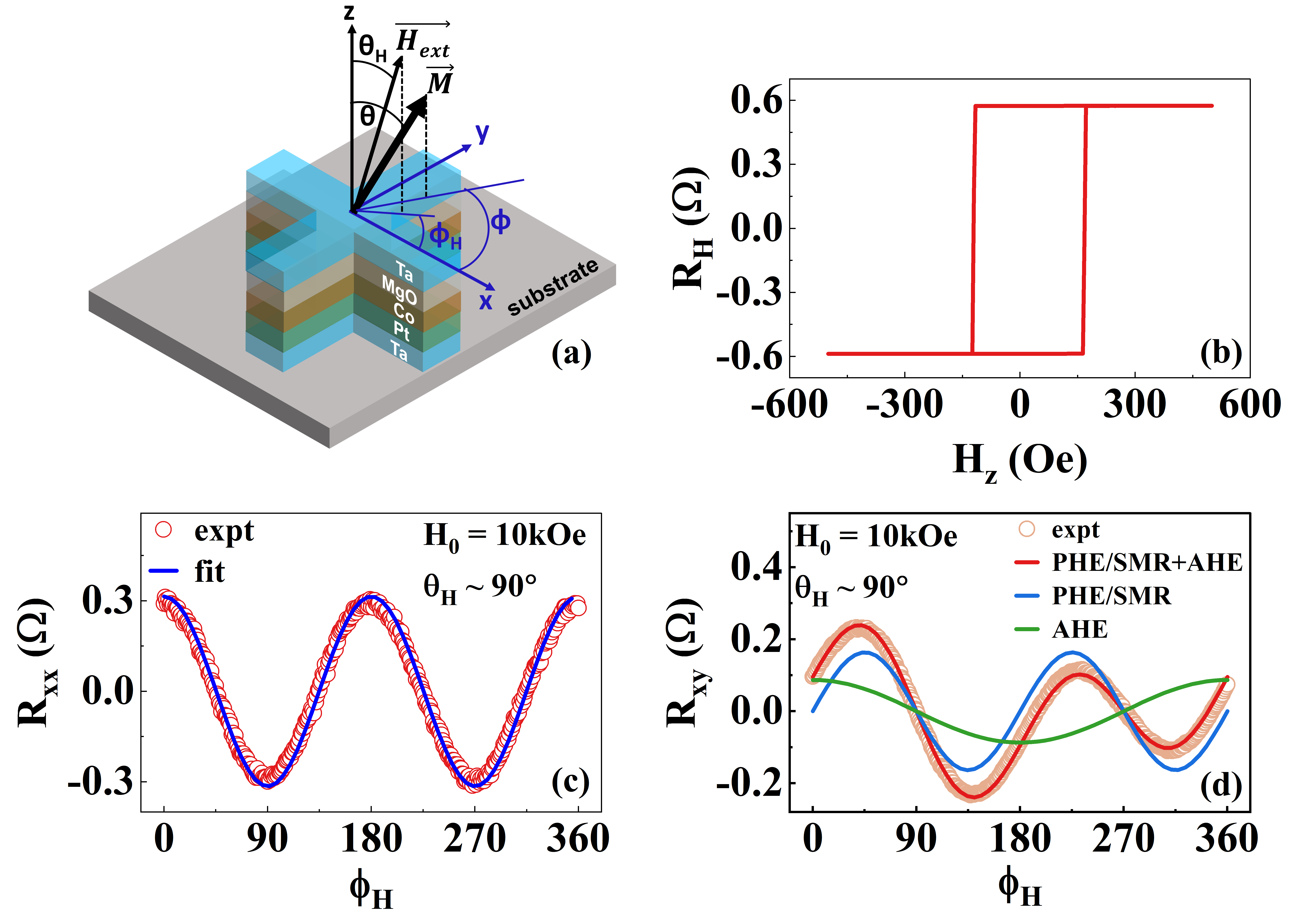}
    \vspace{-0.1in}
    \caption{(a) Schematic of the Pt/Co PMA stack Hall bar along with the reference frame (b) AHE hysteresis loop of the Pt/Co PMA stack (c)(d) Longitudinal and transverse resistance plots respectively of Pt/Co PMA stack at an in-plane external field of 10 kOe.}
    \label{fig1}
    \vspace{-0.2in}
\end{figure}

\vspace{1\baselineskip}
\noindent \emph{Pt/Co PMA stack} - Figure \ref{fig1}a shows the Pt/Co PMA stack. The top MgO/Ta layers provide capping to the Pt/Co interfacial perpendicular anisotropy bilayer which is grown on the Ta seed layer. In the reference frame shown in figure  \ref{fig1}a, $\vec{H_{ext}}$ is the external dc magnetic field applied at a polar angle of $\theta_H$ and azimuthal angle of $\phi_H$, whose magnitude is $H_0$. Polar angle $\theta$ and azimuthal angle $\phi$, give the direction of equilibrium magnetization ($\vec{M}$). As magnetization oscillation caused by SOT, itself causes modulation of AHE and PHE resistances, therefore, first we quantify the AHE and PHE resistances and then proceed to harmonic Hall measurements. The AHE measurement is shown in figure  \ref{fig1}b where the Hall resistance ($R_H$) is plotted as a function of applied out-of-plane magnetic field ($H_Z$). The $\Delta R$ of the $R_H - H_z$ plot is denoted by $\Delta R_A$ and is found to be 1.15 $\Omega$ for the Pt/Co PMA stack. The out-of-plane switching field is 150 Oe. Longitudinal and transverse resistance experimental plots shown in figure \ref{fig1}c and figure \ref{fig1}d respectively, were obtained by in-plane rotation ($\theta_H=90^\circ$) of 10kOe external magnetic field. The 0.6 ohm $\Delta R$ obtained in the longitudinal $R_{xx}$ vs $\phi_H$ plot is the $\Delta R$ of anisotropic magnetoresistance. The transverse $R_{xy}$ vs $\phi_H$ plot in figure \ref{fig1}d has uneven waveshape as this has contributions from both PHE/SMR and AHE. The blue and green curves show the contributions of PHE/SMR ($\Delta R_P$) and AHE respectively. The $R_{xy}$ experimental data was fitted assuming a small tilt of the rotation axis from the $\hat{z}$ direction. The red curve shows the fit obtained with tilt of $2.7^\circ$ for a ($\Delta R_P$) of 0.3$\Omega$. The $\Delta R_A$ obtained from figure \ref{fig1}b and $\Delta R_P$ obtained from figure \ref{fig1}d will be used to fit the harmonic Hall signals and obtain anisotropy field ($H_{ani}$), $\xi_{DL}$ and $\xi_{FL}$.


\begin{figure}[!t]
   \centering
   \vspace{0.0in}
    \includegraphics[width=3in, keepaspectratio]{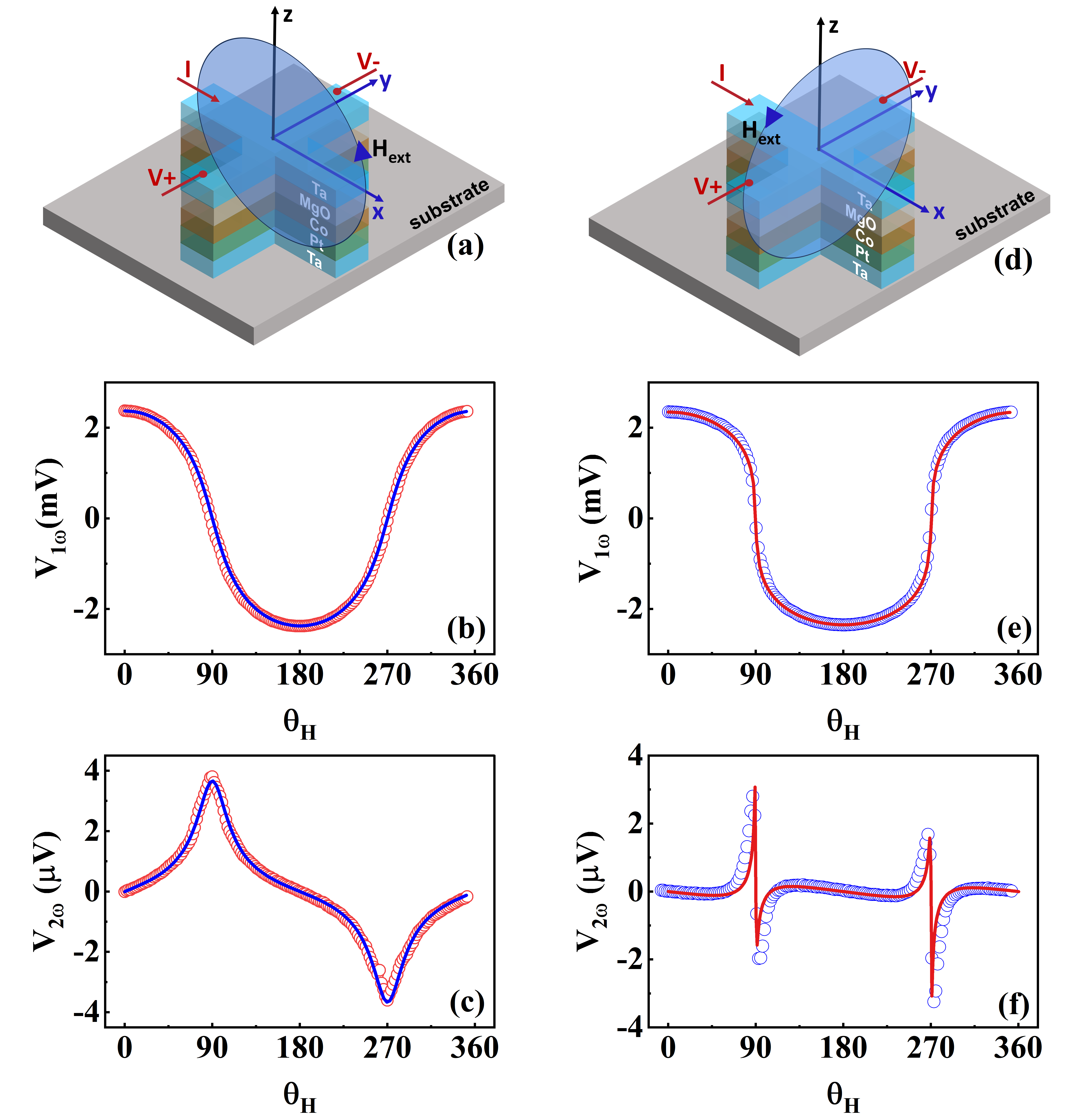}
    \vspace{-0.1in}
    \caption{Out-of-plane angle sweep harmonic Hall measurements of Pt/Co PMA stack at 4mA current (symbols represent experimental data and solid lines represent fit to the data). (a) Schematic of OOP rotation of external magnetic field direction in xz plane for longitudinal case (b)-(c) First and second harmonic Hall voltages for longitudinal case at 20 kOe (d) Schematic of OOP rotation in yz plane for transverse case (e)-(f) First and second harmonic Hall voltages for transverse case at 10 kOe }
    \label{fig2}
    \vspace{-0.2in}
\end{figure}

Next, we present the OOP angle resolved harmonic Hall measurements for the Pt/Co PMA system. In this technique, an out-of-plane $360^\circ$ polar angle ($\theta_H$) sweep of the external magnetic field is done and first and second harmonic Hall voltages are recorded. The input current and output voltage are transverse to each other as shown in figure \ref{fig2}a and \ref{fig2}d. There are two cases - longitudinal and transverse, based on rotation of the external magnetic field. In the longitudinal case in figure  \ref{fig2}a, the ac current is applied in $\hat{x}$ direction and a constant external field is rotated in the x-z plane ($\phi_H = 0^{\circ}$). In the transverse case in figure  \ref{fig2}d, the ac current is still in $\hat{x}$ direction and the constant external field rotation is in the y-z plane ($\phi_H = 90^{\circ}$). In both the cases, the Hall measurements were carried out at a low frequency of 1277 Hz obtained through a lockin amplifier. The low frequency ac current of 4mA was passed through one arm of Hall bar and transverse voltage was read across the other arm of the Hall bar. Figure  \ref{fig2}b and \ref{fig2}c, show the first and second harmonic Hall voltages obtained for the longitudinal case at a polar rotational field of 20 kOe. Figure  \ref{fig2}e and \ref{fig2}f, show the $V_{1\omega}$ and $V_{2\omega}$ obtained for the transverse case at a polar rotational field of 10 kOe. The formalism to fit $V_{1\omega}$ and $V_{2\omega}$ and estimate SOT efficiencies by using low frequency limit of magnetic susceptibility is derived in the next section (Harmonic Hall Analysis: Formalism). An anisotropy field ($H_{ani}$) of 9400 Oe was obtained by individually fitting the $V_{1\omega}$ data in figure \ref{fig2}b and \ref{fig2}e to first term of equation \ref{eq_voltage2}. The $V_{2\omega}$ of the longitudinal case reveals the damping-like SOT efficiency ($\xi_{DL}$) which was obtained by fitting the $V_{2\omega}$ data  to equation \ref{eq_Vdc_long}. The $V_{2\omega}$ of the transverse case reveals the field-like SOT efficiency ($\xi_{FL}$) which was obtained by fitting the $V_{2\omega}$ data to equation \ref{eq_Vdc_trans}. For an anisotropy field of 9400 Oe, $\Delta R_A$ of 1.15 $\Omega$ and $\Delta R_P$ of 0.3 $\Omega$, $\xi_{DL}$ equal to 0.12 and $\xi_{FL}$ equal to -0.05 was obtained.




\begin{figure}[!b]
   \centering
   \vspace{0.0in}
    \includegraphics[width=3in, keepaspectratio]{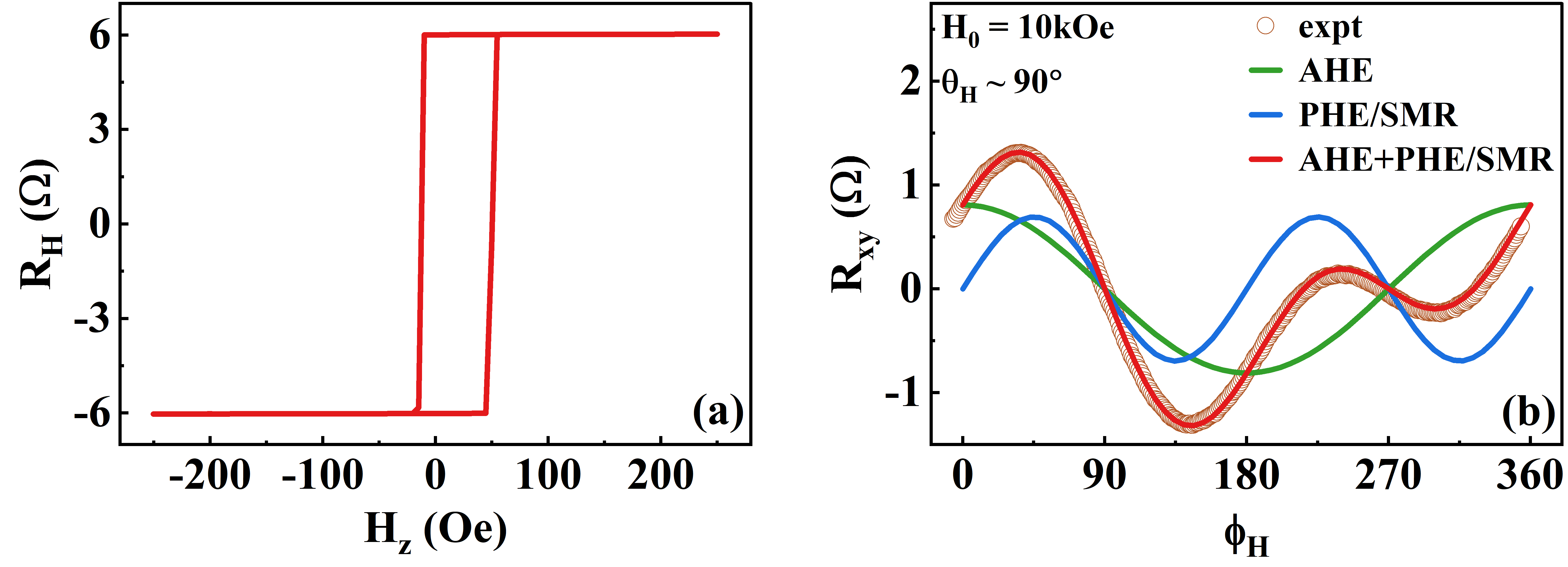}
    \vspace{-0.1in}
    \caption{(a) AHE hysteresis loop of the Ta/CoFeB PMA stack (b) $R_{xy}$ vs $\phi_H$ plot showing contributions from PHE/SMR and AHE}
    \label{fig3}
\end{figure}

\begin{figure}[!t]
   \centering
   \vspace{0.0in}
    \includegraphics[width=3in, keepaspectratio]{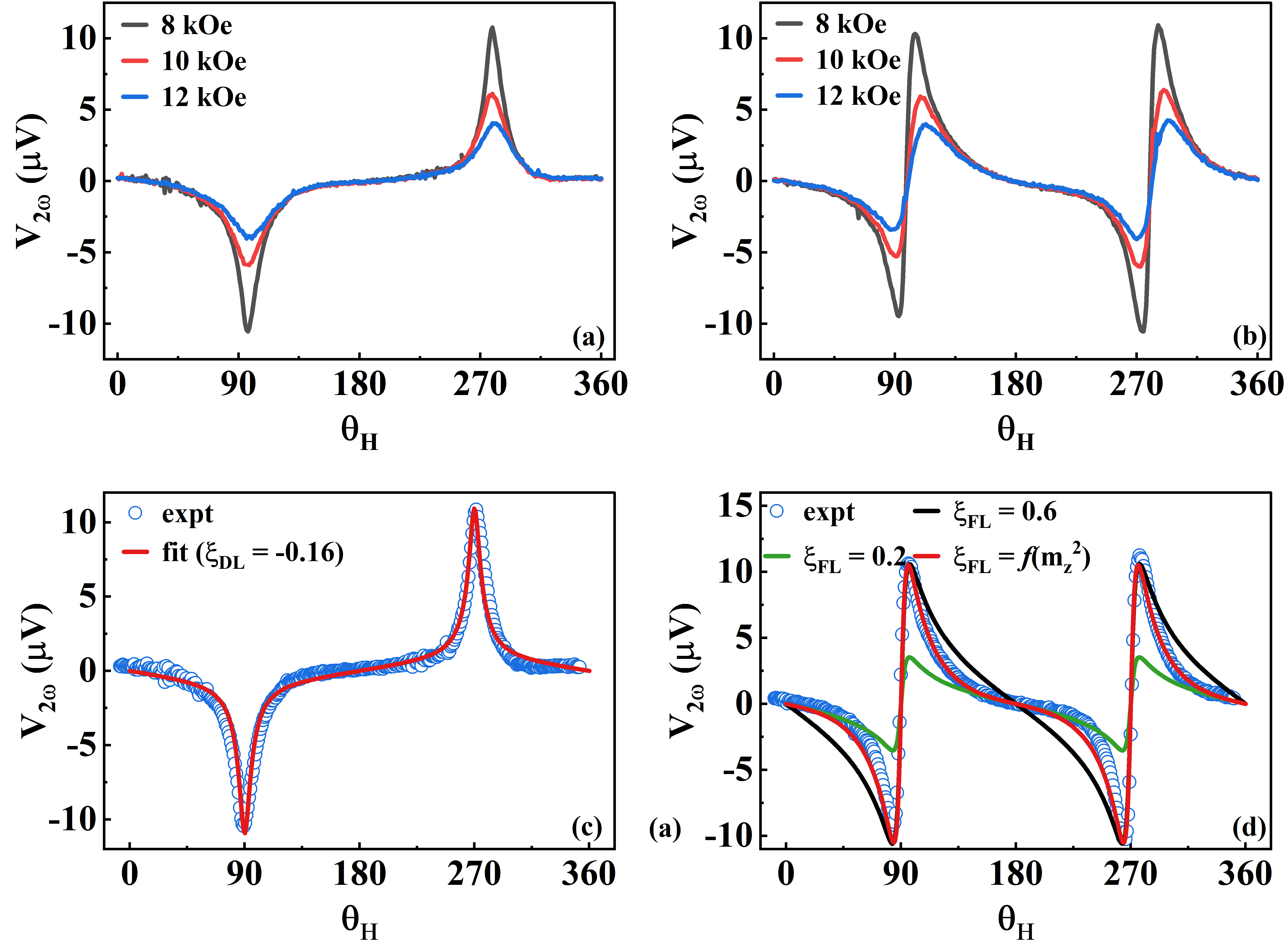}
    \vspace{-0.1in}
    \caption{SHH plots of PMA Ta/CoFeB system acquired at 0.6mA current by OOP angle sweep (a)-(b) $V_{2w}$ for longitudinal and transverse scans respectively for 8, 10 and 12 kOe externally applied magnetic fields. (c) SHH plot at 8000 Oe in longitudinal case (d) SHH plot at 8000 Oe in transverse case}
    \label{fig4}
    \vspace{-0.1in}
\end{figure}

\vspace{1\baselineskip}
\noindent \emph{Ta/CoFeB PMA stack} - The deposited stack for the Ta/CoFeB PMA system is Ta(5)/CoFeB(1.3)/MgO(1.5)/Ta(1.5). Ta is the heavy metal providing spin current to the ferromagnetic CoFeB. The PMA in this system originates at the CoFeB/MgO interface. Using the same methods as described in the previous Pt/Co section, $\Delta R_A$, $\Delta R_P$, anisotropy field and SOT efficiencies were obtained for the Ta/CoFeB PMA stack. Figure \ref{fig3}a shows the AHE $R_H$ vs $H_z$ measurement plot from which $\Delta R_A$ of 12$\Omega$ was obtained for Ta/CoFeB stack. In figure \ref{fig3}b, transverse resistance is plotted w.r.t in-plane angle where the external magnetic field was rotated in the plane of the sample. Like in figure \ref{fig1}d here also, the $R_{xy}$ has contributions from PHE/SMR and AHE. The PHE/SMR contribution denoted by $\Delta R_P$ is found to be 1$\Omega$. 

Next, the OOP angle sweep harmonic Hall measurements for the Ta/CoFeB PMA system were done for both longitudinal and transverse case. The $V_{1\omega}$  (not shown here) for Ta/CoFeB stack has the same nature as the $V_{1\omega}$ obtained in Pt/Co stack. From the $V_{1\omega}$ signal recorded w.r.t to the polar angle, an anisotropy field of 6000 Oe was obtained for Ta/CoFeB stack. Figure \ref{fig4}a and \ref{fig4}b show the plots of SHH voltage in longitudinal and transverse case respectively, at various applied magnetic field values. The fit to the longitudinal data (figure \ref{fig4}c) using equation \ref{eq_Vdc_long}, provides a $\xi_{DL}$ of -0.16. The fit to the transverse data (figure \ref{fig4}d) to estimate $\xi_{FL}$ using equation \ref{eq_Vdc_trans} isn’t as straightforward. We observe that the $\xi_{FL}$ value that fits the data region around $\theta=90^\circ$, does not fit the slope around $180^\circ$. The $\theta=90^\circ$ region fits to a $\xi_{FL}$ of 0.6 and the slope in the vicinity of $180^\circ$ fits to a $\xi_{FL}$ of 0.2. Such large values of $\xi_{FL} > \xi_{DL}$ have been reported in other systems as well \cite{pai2014enhancement}. A better fit to the signal over the whole $360^\circ$ range is obtained by modifying the $h_{FL}$ effective field to $h_{FL}\times(1 + a\, cos^2 \theta)$, where $a=-0.66$. Thus the $h_{FL}$ (and $\xi_{FL}$) is a function of $m_z^2$ in the Ta/CoFeB system. Such anomalous behavior was not observed in the Pt/Co system. It should be noted that, for an in-plane angle sweep of the same Ta/CoFeB device a $\xi_{FL}$ of 0.2 was obtained (data not shown).  The fit to the longitudinal data shown in figure \ref{fig4}c depends on $\xi_{FL}$, but is not very sensitive to it. Thus the the transverse angle scan is essential to uncover such anomalous behavior. The $h_{DL}$ remains odd under time reversal and even under space reversal due to its $m_z^2$  dependence. It should be noted that the observed $m_z^2$ dependence of field like term is allowed by the symmetry of the system. 

We now discuss some important features of the data obtained by OOP angle resolved longitudinal and transverse scans for Pt/Co and Ta/CoFeB. For both the stacks, in the longitudinal scan, the signal shows a maxima/minima at $\theta=90^\circ, 270^\circ$, whose value (and sign) is mainly determined by the damping-like term.  Further, the slopes at $\theta=0^\circ$ and $180^\circ$ are opposite (see equations \ref{eq_Vdc_long_slope} and \ref{eq_Vdc_long_max}). In the case of transverse scan, the slopes at $\theta=0^\circ$ and $180^\circ$ are equal. Further the signal at $\theta=90^\circ, 270^\circ$ is zero, and the slopes at these values are equal. However these slopes at $90^\circ$ and $270^\circ$ are larger than slopes at $0^\circ$ and $180^\circ$, and are mainly determined by  field-like term. It is important to note that the transverse data for Pt/Co and Ta/CoFeB have different features. In the case of Ta/CoFeB, the slopes at $\theta=90^\circ$ and $180^\circ$ are opposite sign (see figure \ref{fig4}d) where as they are of the same sign in case of Pt/Co (see figure \ref{fig2}f). This happens due to the negative field-like term in Pt/Co. Also as shown earlier, to fit the transverse $V_{2\omega}$ fully in Ta/CoFeB system a modified $h_{FL}$ is required whereas the Pt/Co transverse $V_{2\omega}$ fits fully over the $360^\circ$ polar angle range without any modification in the $h_{FL}$ effective field.

\begin{figure}[!t]
   \centering
   \vspace{0.0in}
    \includegraphics[width=3in, keepaspectratio]{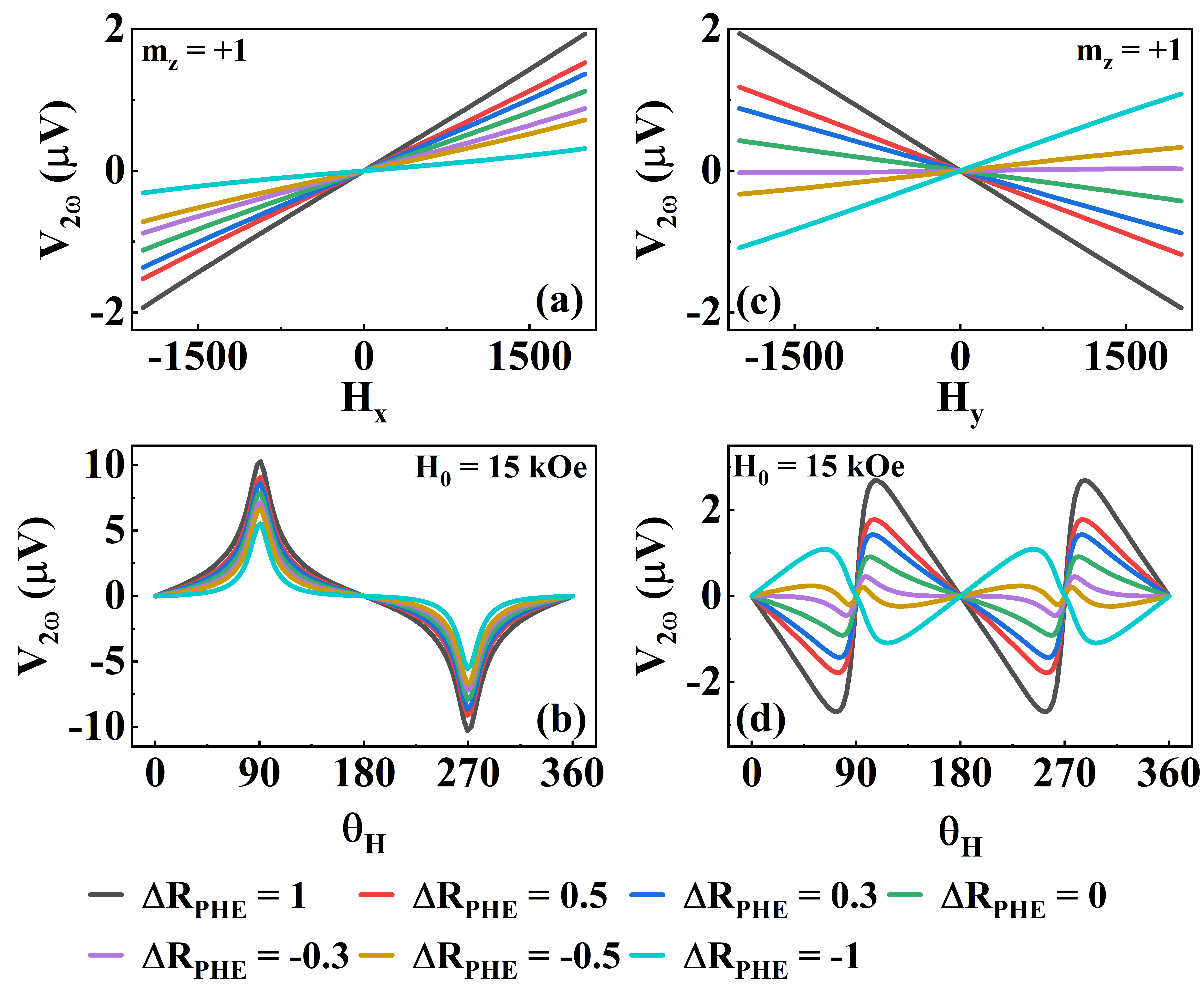}
    \vspace{-0.1in}
    \caption{OOP angle sweep $V_{2\omega}$ and $V_{dc}$ signals of Pt/Co PMA stack at 10 kOe externally applied magnetic field for (a) longitudinal case (b) transverse case }
    \label{fig5}
    \vspace{-0.1in}
\end{figure}

\vspace{1\baselineskip}
\noindent \emph{Equivalence of SHH and dc voltage} - In figure \ref{fig5}, OOP angle sweep $V_{2\omega}$ and $V_{dc}$ plots of Pt/Co PMA stack are shown at an externally applied magnetic field of 10 kOe for longitudinal as well as transverse case. The ac input current to the Hall bar device was given through lockin amplifier. In one case the second harmonic Hall voltage ($V_{2\omega}$) was recorded through lockin amplifier and in other case the Hall voltage was recorded through nanovoltmeter which measured $V_{dc}$. From figure \ref{fig5}, it can be observed that the $V_{2\omega}$ and $-V_{dc}$ signals overlap and are equivalent. For an ac input current ($I_x sin \omega t$), the time evolution of the Hall voltage is proportional to $sin^2(\omega t)$. This $sin^2(\omega t)$ results in a dc voltage term and a second harmonic voltage term whose magnitude is same as the dc voltage (see equation \ref{eq_equivalence} in the Harmonic Hall Analysis: Formalism section).

\section{Harmonic Hall Analysis: Formalism}

The reference frame used in the analysis is shown in figure \ref{fig1}a. The ferromagnetic film is in x-y plane. The external dc magnetic field is applied at polar angle of $\theta_H$ and azimuthal angle of $\phi_H$ . The equilibrium magnetization direction is denoted by polar angle of $\theta$ and azimuthal angle of $\phi$.
The free energy density of the FM is taken as: $f=\mu_0 M_s [-\hat{m}.\vec{H_{ext}}-(1/2)H_{\perp}m_z^2] $, where $\hat{m}$ denotes unit vector along magnetization direction, $M_s$ denotes the saturation magnetization and $H_{\perp}$ denotes the effective anisotropy field along $\hat{z}$ direction. The magnetic field acting on the FM is $\vec{H}=-(1/\mu_0)(\partial{F}/\partial{\hat{m}})=\vec{H_{ext}}+\vec{H_{ani}}$ with $\vec{H_{ani}}=H_{\perp}m_z\hat{m}$. The magnetization dynamics including the effect of charge current passing through the heavy metal is given by LLGS equation:
\begin{align}
\label{eq_LLGS1}
&\frac{d\hat{m}}{dt}=-\gamma_0(\hat{m} \times \vec{H_{eff}})+\alpha(\hat{m} \times \frac{d\hat{m}}{dt}) \\
&+\frac{\hbar}{2q}\frac{\gamma_0}{M_s\mu_0 t_{FM}} J_{HM}[\xi_{DL}(\hat{m}\times (\hat{y} \times \hat{m}) )-\xi_{FL} (\hat{m}\times \hat{y}) ] && \nonumber
\end{align}
where $\vec{H_{eff}}=\vec{H_{ext}}+\vec{H_{ani}}+\vec{H_{Oe}}$. $\vec{H_{Oe}}=\hat{y}J_{HM}t_{HM}/2$ is the Oersted magnetic field produced by the current.  $J_{HM}$ is the charge current density through heavy metal, $t_{HM}$ and $t_{FM}$ are the thickness of heavy metal and ferromagnet respectively and $\alpha$ is the damping coefficient. Defining effective magnetic fields produced by the spin current as: 
$h_{DL}=\frac{\hbar}{2q}\frac{1}{M_s\mu_0 t_{FM}} J_{HM}\xi_{DL}$ and $h_{FL}=\frac{\hbar}{2q}\frac{1}{M_s\mu_0 t_{FM}} J_{HM}\xi_{FL}$, the above equation \ref{eq_LLGS1} can be written as
\begin{align}
\label{eq_LLGS2}
&\frac{d\hat{m}}{dt}=-\gamma_0(\hat{m} \times \vec{H_{eff}})+\alpha(\hat{m} \times \frac{d\hat{m}}{dt}) &\\
&+\gamma_0[(\hat{m}\times ( \vec{h_{DL}} \times \hat{m}) )- (\hat{m}\times \vec{h_{FL}}) ] && \nonumber
\end{align}
where $\vec{h_{DL}}=h_{DL}\hat{y}$ and $\vec{h_{FL}}=h_{FL}\hat{y}$. 
Note that the above equations are written for sub/FM/HM stack. For sub/HM/FM stack the expressions for $\vec{H_{Oe}}$, $\vec{h_{FL}}$, and $\vec{h_{DL}}$ should be multiplied by $-1$. 

The equilibrium magnetization direction is taken as x' axis. i.e. $\hat{x'}$ $=[sin(\theta)cos(\phi),sin(\theta)sin(\phi),cos(\theta)]$. 
The y' axis is chosen to lie in the z-x' plane, and thus
the y' and z' directions are taken as: $\hat{y'}=[cos(\theta)cos(\phi),cos(\theta)sin(\phi),-sin(\theta)]$ and $\hat{z'}=[-sin(\phi),cos(\phi),0]$. Note that since we have considered only uni-axial anisotropy along z-axis, $\phi=\phi_H$.
If a small magnetic field and damping-like magnetic field oscillating with angular frequency $\omega$ are applied to the FM, the oscillating magnetization in the dash frame can be written as:
\begin{gather}
\label{eq_for_delta_m}
 \begin{bmatrix} \delta m_{x'}  \\ \delta m_{y'} \\ \delta m_{z'} \end{bmatrix}
 =
   \begin{bmatrix}
   0 & 0 & 0 \\
   0 & \chi_{11}^H & \chi_{12}^H \\
   0 & \chi_{21}^H & \chi_{22}^H  
   \end{bmatrix}
   \begin{bmatrix} h_{x'}  \\ h_{y'} \\ h_{z'} \end{bmatrix}
   +\begin{bmatrix}
   0 & 0 & 0 \\
   0 & \chi_{11}^I & \chi_{12}^I \\
   0 & \chi_{21}^I & \chi_{22}^I  
   \end{bmatrix}
   \begin{bmatrix} h_{DL,x'}  \\ h_{DL,y'} \\ h_{DL,z'} \end{bmatrix}
\end{gather}
where $\chi^H$ denotes the Polder susceptibility calculated in the dash frame, given by:
\begin{gather}
\label{eq_sus1}
\begin{bmatrix} \chi_{11}^H & \chi_{12}^H \\ \chi_{21}^H & \chi_{22}^H \end{bmatrix}
 =\frac{\gamma}{-\omega^2+(1+\alpha^2)\omega_1\omega_2-i\alpha\omega(\omega_1+\omega_2)} \\
   \begin{bmatrix} -i\alpha\omega+(1+\alpha^2)\omega_1 & -i\omega 
   \\ i\omega &-i\alpha\omega+(1+\alpha^2)\omega_2  \nonumber
   \end{bmatrix}
\end{gather}
where $\gamma=\gamma_0/(1+\alpha^2)$, $H_0$ denotes the magnitude of the external magnetic field and
\begin{align}
\label{eq_omega1}
\omega_1=\gamma[H_{\perp}cos^2\theta +H_0 cos(\theta-\theta_H)] \\
\omega_2=\gamma[H_{\perp}cos(2\theta)+H_0 cos(\theta-\theta_H)]
\label{eq_omega2}
\end{align}

The susceptibility due to the damping-like field is given by:
\begin{gather}
 \begin{bmatrix} \chi_{11}^I & \chi_{12}^I \\ \chi_{21}^I & \chi_{22}^I \end{bmatrix}
=\begin{bmatrix} \chi_{12}^H & -\chi_{11}^H \\ \chi_{22}^H & -\chi_{21}^H \end{bmatrix}
\label{eqn_sus2}
\end{gather}

The oscillating magnetization in the (x,y,z) frame can be written as:
\begin{gather}
 \label{eq_for_delta_m22}
 \begin{bmatrix} \delta m_{x}  \\ \delta m_{y} \\ \delta m_{z} \end{bmatrix}
 =R
   \begin{bmatrix}
   0 & 0 & 0 \\
   0 & \chi_{11}^H & \chi_{12}^H \\
   0 & \chi_{21}^H & \chi_{22}^H  
   \end{bmatrix} 
   R^T 
   \begin{bmatrix} h_{x}  \\ h_{y} \\ h_{z} \end{bmatrix} \\ \nonumber
   + R
   \begin{bmatrix}
   0 & 0 & 0 \\
   0 & \chi_{11}^I & \chi_{12}^I \\
   0 & \chi_{21}^I & \chi_{22}^I  
   \end{bmatrix}
   R^T
   \begin{bmatrix} h_{DL,x}  \\ h_{DL,y} \\ h_{DL,z} \end{bmatrix}
\end{gather}
where R is the rotation matrix connecting the two reference frames, given by:
\begin{gather}
 R=\begin{bmatrix} 
 cos\phi \, sin\theta & cos\phi \, cos\theta & -sin\phi   
 \\ sin\phi \, sin\theta & cos\theta \, sin\phi & cos\phi  
 \\ cos\theta & -sin\theta & 0
 \end{bmatrix}
 \label{eq_R_mat}
\end{gather}
We can easily obtain expressions for the oscillating magnetization using equation  \ref{eq_for_delta_m22} and \ref{eq_R_mat} for e.g. $\delta m_z$ arising from oscillating magnetic field is given by:
\begin{align}
&\delta m_{z}=sin\theta (\chi_{12}^H sin\phi -\chi_{11}^H cos\theta \, \cos\phi )h_x \\
&-sin\theta (\chi_{12}^H cos\phi +\chi_{11}^H cos\theta \, sin\phi )h_y 
+ \chi_{11}^H sin^2\theta h_z \nonumber
\end{align}
Similar expression can be written for the contribution of the oscillating damping-like magnetic field, by replacing $\chi^H$ with $\chi^I$ and $h$ by $h_{DL}$.

\vspace{1\baselineskip}
\noindent\textbf{Calculation of the second harmonic Hall and dc voltage}: We pass a small ac current along $\hat{x}$ direction and measure the second harmonic or dc voltage along y direction. The transverse resistance is taken as:
\begin{equation}
R_{xy}=(\Delta R_A/2) m_z + \Delta R_P m_xm_y
\label{eq_deltaR}
\end{equation}
where the first term in the above equation arises from the anomalous Hall effect, and the second term from planar Hall effect. The ac current which passes through the HM layer, exerts a torque on the magnetization as given by equation  \ref{eq_LLGS2}. This gives rise to the oscillation of magnetization as given by equation \ref{eq_for_delta_m22}, which in turn results in the oscillation of the transverse resistance. The transverse voltage resulting from the combination of time dependent resistance and  current is given by equation \ref{eq_voltage1} below. Note that the contributions from spin-pumping and magnetic induction to the transverse voltage are not included in equation \ref{eq_voltage1}.
\begin{align}
\label{eq_voltage1}
&V_y(t)=[(\Delta R_A/2) (m_{z0}+\delta m_z(t)) \\
&+ \Delta R_P (m_{x0}+\delta m_x(t))(m_{y0}+\delta m_y(t))] I_{x} cos(\omega t) \nonumber
\end{align}
where $m_{x0}$, $m_{y0}$ and $m_{z0}$ denote the components of equilibrium magnetization direction and are given by $sin\theta cos\phi$, $sin\theta sin\phi$ and $cos\theta$ respectively. In equation \ref{eq_voltage1}, we can put $\delta m_z(t)=\delta m_z exp(-i\omega t)=Re(\delta m_z)cos(\omega t)-Im(\delta m_z)sin(\omega t)$ and similar expressions for $\delta m_x(t)$ and $\delta m_y(t)$.  The transverse voltage is given by equation \ref{eq_voltage2} below.

\begin{align}
\label{eq_voltage2}
&V_y(t) = [(\Delta R_A/2) m_{z0} + \Delta R_P m_{x0} m_{y0}] I_{x} cos(\omega t) \\
&+ (\frac{\Delta R_A}{2})[Re(\delta m_z) cos^2\omega t-Im(\delta m_z) sin\omega t \,cos\omega t]I_{x} && \nonumber \\
&+\Delta R_P m_{x0}[Re(\delta m_y) cos^2\omega t-Im(\delta m_y) sin\omega t \,cos\omega t]I_{x} && \nonumber \\
&+\Delta R_P m_{y0}[Re(\delta m_x) cos^2\omega t-Im(\delta m_x) sin\omega t \,cos\omega t]I_{x} && \nonumber
\end{align}

The above equation \ref{eq_voltage2} can be used to get the dc voltage and the second harmonic response.  If we consider low frequency excitation, the susceptibility is given by (see equation \ref{eq_sus1}):
\begin{gather}
 \label{eq_sus3}
 \begin{bmatrix} \chi_{11}^H & \chi_{12}^H \\ \chi{21}^H & \chi_{22}^H \end{bmatrix}
 \approx \gamma \begin{bmatrix} \frac{1}{\omega_2} & 0 \\ 0 & \frac{1}{\omega_1}  \end{bmatrix} \\ \nonumber
 =\begin{bmatrix} \frac{1}{H_{\perp}cos(2\theta)+H_0 cos(\theta-\theta_H)} & 0 \\ 0 & \frac{1}{H_{\perp}cos^2(\theta)+H_0 cos(\theta-\theta_H)}  \end{bmatrix}
\end{gather}
The low frequency susceptibility is independent of $\alpha$ and frequency and is real. Thus all the terms involving imaginary part in equation \ref{eq_voltage2} are absent. Using, $cos^2(\omega t)=(1+cos(2\omega t))/2$, the dc voltage is given by equation \ref{eq_Vdc1} below.
\begin{align}
\label{eq_Vdc1}
&V_{dc}=\{ (\Delta R_A/4) Re(\delta m_z) \\ \nonumber
&+ (\Delta R_P/2) [m_{x0}Re(\delta m_y)+ m_{y0} Re(\delta m_x)]\}I_{x}
\end{align}
For measurement of second harmonic response using lock-in amplifier, the excitation is taken as $sin(\omega t)$.  Using, $sin^2(\omega t)=(1-cos(2\omega t))/2$, we get, 

\begin{equation}
\label{eq_equivalence}
    V_{2\omega}=-V_{dc}
\end{equation}

In this experiment, all the oscillating fields are along $\hat{y}$ direction and the expression for $V_{dc}$ can be easily obtained using the above set of equations. The general expressions  obtained are given below:
\begin{align}
\label{eq_Vdc_final1}
& V_{dc,AHE}=\frac{\Delta R_A}{4} \chi_{11}^H sin\theta(-cos\theta \, sin\phi h_y+cos\phi h_{DL})  I_{x} \\ 
\label{eq_Vdc_final2}
& V_{dc,PHE,DL}=-(\frac{\Delta R_P}{2}) cos\theta \, sin\theta \, sin\phi \\ \nonumber 
&[ \chi_{11}^H +cos(2\phi)(\chi_{11}^H-\chi_{22}^H )]h_{DL} I_{x} \\ 
\label{eq_Vdc_final3}
& V_{dc,PHE,FL}=(\frac{\Delta R_P}{2})  sin\theta \, cos\phi \\ \nonumber
\label{eq_Vdc_final4}
&[2\chi_{11}^H sin^2\phi \,cos^2\theta+\chi_{22}^H \, cos(2\phi) ] h_y I_{x} \\ 
& V_{dc}=V_{dc,AHE}+V_{dc,PHE,DL}+V_{dc,PHE,FL} 
\end{align}
where $h_y=h_{FL}+H_{Oe}$. 
We have compared the expression of $V_{dc,AHE}$ given by equation  \ref{eq_Vdc_final1},  with obtained by Hayashi et. al. \cite{hayashi2014quantitative}. Putting, $\Delta H_x=-cos\theta h_{DL}$, $\Delta H_z=sin\theta cos\phi h_{DL}$ and $H_A=0$ in equation 14 of their paper,  the expressions of $V_{2\omega}$ match. The experimental measurements have been carried out in two geometries in this work: longitudinal ($\phi=0$ or $ 180^o$) and transverse ($\phi=90^o$ or $-90^o$). For these two cases, the expressions for dc voltage given by equation \ref{eq_Vdc_final1}-\ref{eq_Vdc_final4} are simplified and given below:
\begin{align}
\label{eq_Vdc_long}
& V_{dc}=[\frac{\Delta R_A}{4} \chi_{11}^H h_{DL}+\frac{\Delta R_P}{2} \chi_{22}^H h_{y}]    sin\theta I_{x} \,\,(\phi=0) \\ 
\label{eq_Vdc_trans}
& V_{dc}=-[\frac{\Delta R_A}{4} \chi_{11}^H h_{y}+\frac{\Delta R_P}{2} \chi_{22}^H h_{DL}] sin\theta cos\theta I_{x}   \,\,(\phi=90^o)
\end{align}

The above equations have been used for fitting the data shown in figure \ref{fig2}c, \ref{fig2}f, \ref{fig4}c and \ref{fig4}d. These equations need $\theta$ and $\phi$, (i.e. the equilibrium direction of magnetization), which are obtained from the solution of LLG equation for a given $\vec{H_{ext}}$. It is also useful to look at the derivative of $V_{dc}$ at various $\theta$ values such as $0, 90^o, 180^0$ and $270^o$ which could be used for evaluating  damping-like and field-like terms. We now assume that the external field ($H_0$) is more that the anisotropy field ($H_{ani}$), so that the magnetization points along the external field at these $\theta$ values. These derivatives can be obtained by noting that the low frequency susceptibility at $\theta=0,180^o$,   is given by :  $\chi_{11}^H = \chi_{22}^H = 1/(H_\perp +H_0)$ , where as at $\theta=90^o, 270^o$, $\chi_{11}^H =1/(H_0-H_{\perp})$ and $ \chi_{22}^H = 1/H_0$ (see equation  \ref{eq_sus3}).  In the longitudinal case we find that:
\begin{align}
\label{eq_Vdc_long_slope}
& \frac{dV_{dc}}{d\theta}=\pm[\frac{\Delta R_A}{4} h_{DL}+\frac{\Delta R_P}{2} ]\frac{1}{H_0+H_{\perp}}I_{x} (\theta=0,180^o) \\ 
\label{eq_Vdc_long_max}
& V_{dc}= \pm [\frac{\Delta R_A}{4} \frac{1}{H_0-H_{\perp}}  h_{DL}+\frac{\Delta R_P}{2} \frac{1}{H_0} h_{y}]  I_{x} (\theta=90^o, 270^o)
\end{align}
The value of $V_{dc}$ is maximum (or minimum) at $\theta=90^\circ$ or $270^\circ$.
In the transverse case, we find:
\begin{align}
\label{eq_Vdc_trans_slope}
& \frac{dV_{dc}}{d\theta}=-[\frac{ \Delta R_A}{4} h_{DL}+ \frac{\Delta R_P}{2} h_{y}] \frac{1}{H_0+H_{\perp}}I_{x} (\theta=0^o, 180^o)  \\
\label{eq_Vdc_trans_slope2}
& \frac{dV_{dc}}{d\theta}=[\frac{ \Delta R_A}{4} \frac{1}{H_0-H_{\perp}} h_{DL}+ \frac{\Delta R_P}{2} \frac{1}{H_0} h_{y}] I_{x} (\theta=90^o, 270^o) 
\end{align}


The $V_{1\omega}$ plots in figure \ref{fig2}b and \ref{fig2}e have been fitted using the first term in equation  \ref{eq_voltage2}. Since $m_{y0}$ is zero for longitudinal case and $m_{x0}$ is zero for transverse case, the contribution to $V_{1\omega}$ arises only from the AHE. The $m_{z0}$ was obtained from the solution of LLG equation.

Equation \ref{eq_voltage2} can be used to obtain the dc voltage measured in STFMR experiment. We now consider a case where external magnetic field more than the anisotropy field of PMA ferromagnet is applied in the xy plane. The magnetization in this case lies  in the xy plane. This geometry has been used for the AHE based STFMR experiment in the next section. The expression for the dc voltage is given in equation \ref{eq_stfmr} below-
\begin{align}
    \label{eq_stfmr}
    & V_{dc}=\frac{\Delta R_A}{4} cos\phi [(-Re(\chi_{12}^H) h +  Re(\chi_{11}^H) h_{DL}] \\
    & +\frac{\Delta R_P}{2}  cos\phi \,cos2\phi [Re(\chi_{22}^H) h -  Re(\chi_{21}^H) h_{DL}] \} \nonumber
\end{align}
where the susceptibility is given by equation \ref{eq_sus1}. We can see that if the angle $\phi$ is $45^o$, the dc voltage has no contribution from the PHE \cite{bose2017sensitive}, and it arises entirely from the AHE. Further, the peak in the dc voltage arises from the field-like term and dispersion arises from the damping-like term. (This is opposite to what happens in AMR based STFMR experiments.)

\section{AHE based STFMR}

\begin{figure}[h]
   \centering
   \vspace{0.3in}
    \includegraphics[width=3.5in, keepaspectratio]{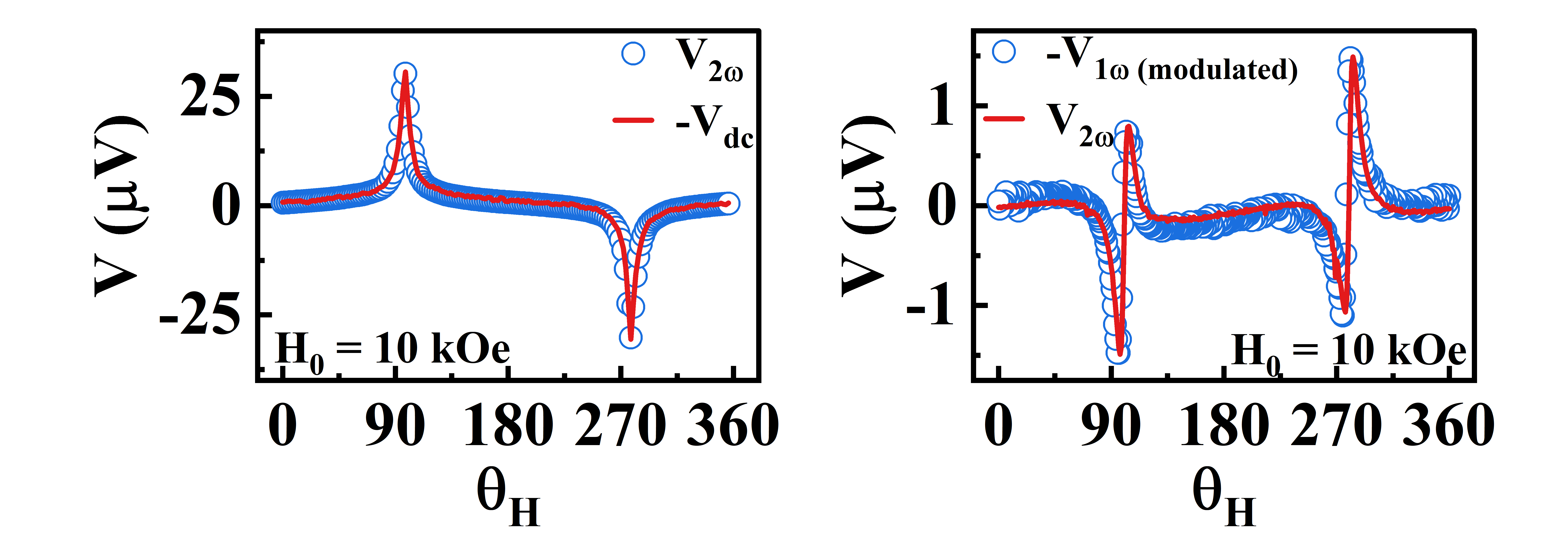}
    \vspace{-0.2in}
    \caption{(a) Measured STFMR signal for multiple frequencies with inset showing the Kittel fit (b) decomposition of STFMR dc voltage signal (above $H_\perp$) into damping-like (DL) and field-like (FL) components }
    \label{fig6}
    \end{figure}

Along with SHH measurement, we performed STFMR measurement also on PMA samples. The AHE based STFMR measurement was performed on 10 $\times$ 20 $\mu$$m^2$ Hall bar devices with a PMA stack of Ta(1)/Pt(5)/Co(1)/MgO(1.5)/Ta(1.5). 
External magnetic field was swept along $\phi=45^o$ direction in xy plane. RF current was flown through the device and transverse dc voltage was read for a magnetic field sweep of $\pm$ 20 kOe. The magnetic field sweep data was recorded for multiple frequencies in the range of 15 to 20 GHz as shown in figure \ref{fig6}a. In general the dc voltage arises from both AHE and PHE. However, as the PHE contribution is proportional to  $cos\phi cos2\phi$ \cite{bose2017sensitive}, the dc voltage arises entirely from the AHE for $\phi=45^o$ angle.

The inset of figure \ref{fig6}a shows the resonant field ($H_{reso}$) vs frequency ($f$) plot. The experimental data is fit to $f=\gamma/2\pi\sqrt{(H_{reso})(H_{reso}-H_\perp)}$. From the fit, the anisotropy field ($H_\perp$) was found to be 5800 Oe. Figure \ref{fig6}b shows the decomposition of the $V_{dc}$ signal into damping-like (DL) and field-like (FL) components, obtained by using equation \ref{eq_stfmr}. The signal has been fitted for magnetic field range which is greater than $H_\perp$, where the magnetization would lie in xy plane. From the fit, $\xi_{DL}$ is estimated to be 0.12 and $\xi_{FL}$ is estimated to be -0.04, which agrees well with SHH results. 

\section{\label{sec:level1}CONCLUSION}
In summary, in this article we demonstrated the out-of-plane angle sweep of the magnetic field approach for harmonic Hall measurement. The SOT efficiencies were estimated for Pt/Co and Ta/CoFeB PMA systems by solving the LLGS equation for PMA systems in the low frequency limit of magnetic susceptibility. The unique variation in $\xi_{FL}$ in Ta/CoFeB PMA system needs to be further studied to understand the mechanisms responsible for such behaviour. Along with this, AHE based STFMR measurements were carried out to estimate the SOT efficiencies.

Authors are thankful for the support provided by IIT-Bombay Nanofabrication Facility, Indian Institute of Technology Bombay.


\bibliography{ref}

\end{document}